# Conductance recovery and spin polarization in boron and nitrogen co-doped graphene nanoribbons

**Seong Sik Kim,[a] Han Seul Kim,[a] Hyo Seok Kim,[a] and Yong-Hoon Kim[a,b,*]**

[a] Graduate School of EEWS, KAIST, 291 Daehak-ro, Yuseong-gu, Daejeon 305-701, Korea.

[b] KI for the NanoCentury, KAIST, 291 Daehak-ro, Yuseong-gu, Daejeon 305-701, Korea.

## Abstract

We present an *ab initio* study of the structural, electronic, and quantum transport properties of B-N-complex edge-doped graphene nanoribbons (GNRs). We find that the B-N edge codoping is energetically a very favorable process and furthermore can achieve novel doping effects that are absent for the single B or N doping. The compensation effect between B and N is predicted to generally recover the excellent electronic transport properties of pristine GNRs. For the zigzag GNRs, however, the spatially localized B-N defect states selectively destroy the doped-side spin-polarized GNR edge currents at the valence and conduction band edges. We show that the energetically and spatially spin-polarized currents survive even in the fully ferromagnetic metallic state and heterojunction configurations. This suggests a simple yet efficient scheme to achieve effectively smooth GNR edges and graphene-based spintronic devices.

[*] Corresponding author. Tel.: +82 42 350 1717. E-mail address: y.h.kim@kaist.ac.kr (Y.-H. Kim)

1. **Introduction**

Adopting the nanoribbon form of graphene [1] is an attractive route toward the next-generation graphene-based electronic [2-5], optoelectronic [6], and spintronic applications [7-11], thanks to the tunable quantum-confined bandgaps [12-15] and spin-polarized edge states [7, 15-18]. For example, room-temperature ballistic charge transport on a length scale greater than ten micrometers was recently demonstrated [19]. To extract such exceptional properties, narrow and clean graphene nanoribbons (GNRs) with atomically smooth edges are required [2, 3, 15, 19-21]. On the other hand, to tune their electronic and charge transport properties for practical device applications, one usually needs to functionalize the GNRs by introducing defects and dopants [22-24]. The challenge is thus to devise a method to controllably functionalize GNRs while maintaining their desirable and unique electronic and transport properties.

In this article, carrying out density functional theory (DFT) and DFT-based non-equilibrium Green's function (NEGF) calculations, we study the structural, electronic, and charge transport properties of B-N-complex edge-doped armchair and zigzag GNRs (aGNRs and zGNRs, respectively). Boron and nitrogen are ideal doping candidates for graphene and GNRs, and significant efforts have been devoted toward the understanding of their physical properties and realization of practical devices based on them [22, 25-31]. However, while theoretical studies have been carried out on single B/N-doped [25-31] or separate B-N co-doped GNRs [30], introducing B-N-complex units into GNRs still remains unexplored. We will show that binary B-N-complex GNR edge doping is energetically a very favorable process, and moreover it almost perfectly recovers the excellent charge transport property of pristine GNRs. For the zGNRs, in particular, we will demonstrate that it can support spatially and energetically spin-polarized currents even without magnetic electrodes and/or external fields. Two heterojunction examples will be provided where the zGNRs are utilized as the spin current channel and spin injection electrodes, respectively.

**2.   Computational method**

We considered both the armchair and zigzag GNRs labeled by the number $N$ of respectively zigzag chains ($N$-aGNRs) and dimer lines ($N$-zGNR) contained in their unit cells. Starting from the pristine GNRs, we introduced single B, N, and B-N-complex dopant atoms into the nanoribbon edges, which were found to be the most favorable doping sites [27, 31] (see Supplemental Data Table S1). In all the cases, pristine or doped GNR edges were terminated by hydrogen atoms. Rectangular simulation boxes with periodic boundary condition were employed, and the inter-GNR distance was kept to a minimum of 20.0 Å to avoid interactions with images.

We carried out DFT calculations within the local density approximations [32] using SIESTA software [33]. The atomic cores were replaced by norm-conserving nonlocal pseudopotentials of Troullier-Martins type [34]. The double $\zeta$-plus-polarization-level numerical atomic orbital basis sets whose extensions were defined by the 80 meV energy cut-off were adopted. The real-space integration was done on the mesh defined by 200 Ry kinetic energy cutoff. The transmission functions of single B, N and B-N-complex doped GNRs were calculated using the fully self-consistent DFT-based NEGF method, as implemented in the TranSIESTA code [35]. The surface Green's functions were extracted from two independent DFT calculations for the two (four) pristine aGNR (zGNR) unit cells that correspond to the electrodes with a 96 $\vec{k}_\perp$-point sampling along the charge transport direction (see Supplemental Data Fig. S1). In calculating the transmission functions T(E), the energy window was scanned from $-1.5$ to 1.5 eV near the Fermi level $E_F$ with the 0.001 eV step. In many cases, using our in-house code [36, 37] and/or the generalized gradient approximation, we have crosschecked the validity of our results (see Supplemental Data Fig. S2). We emphasize that the edge N doping has been experimentally achieved already [21].

**3.   Results and Discussion**

We start with discussing the energetics of B-N-complex doping. Figure 1 summarizes the formation energies of the six-unit-cell-long (length $L$ = 25.56 Å) 11-aGNRs (width $W$ = 14.21 Å) and 12-unit-cell-long ($L$ = 29.52 Å) 6-zGNRs ($W$ = 13.58 Å) doped by binary B and N atoms at one GNR edge in various relative configurations, compared with those of the energetically most stable single-N and single-B doping configurations (edge and next-edge sites, respectively. See also Supplemental Data Fig. S1 and Table S1 for other doping configurations including the GNR interior sites). The formation energies of the B-N doped GNRs were computed using the following equation [28]:

$$E_f = E_{doped} - E_{pristine} + 2\mu_C - \mu_B - \mu_N, \tag{1}$$

where $E_{doped}$ and $E_{pristine}$ are the total energies of the doped/pristine GNRs, and $\mu_{C/B/N}$ are the chemical potential energies of an atomic species C/B/N, respectively. The chemical potentials of C, B, and N were extracted from graphene, $\alpha$-phase bulk B, and N$_2$ gas, respectively. For both aGNR and zGNR cases, we found that the energetically most stable atomic configurations are those where the B and N are chemically bonded to each other and form a complex at the GNR edges. In particular, we note that B-N-complex edge doping is energetically even more favorable than single B or N doping in both aGNRs and zGNRs.

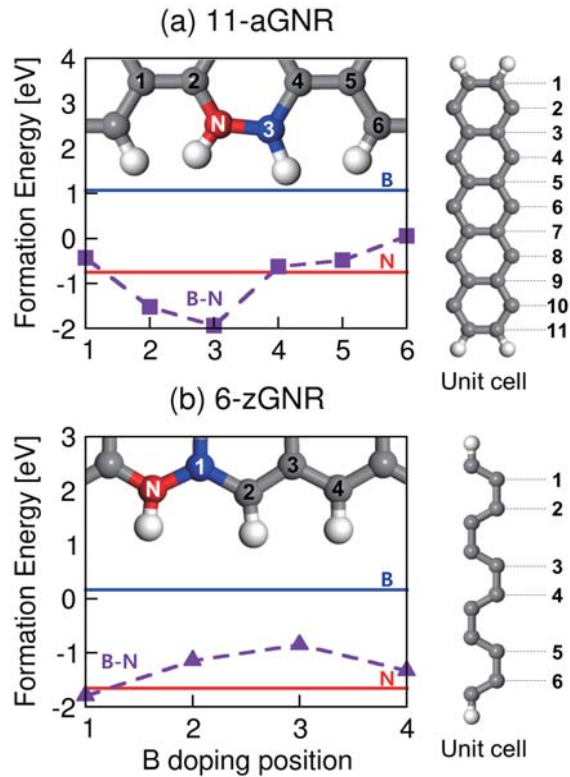

**Fig. 1** Formation energies of single B-, single N- and binary B-N-doped (a) 11-aGNRs and (b) 6-zGNRs. Insets: Doping sites correspond to the B doping positions. The atomic models on the right-hand side represent the unit cell of the 11-aGNR and 6-zGNR respectively.

We now consider the electronic and charge transport properties of the energetically most favorable B-N-complex edge-doped aGNRs. As references, we first show in Figs. 2a and 2b (see also Supplemental Data Fig. S3) the transmissions of single B and single N edge-doped 11-aGNRs, respectively. The data obtained for the pristine 11-aGNR are also shown together as dotted lines. In agreement with previous studies [29, 31], we observe that transmissions are significantly reduced from the pristine GNR values of 1 $G_0$ (= $e^2/h$) in the valence/conduction band regions (p-/n-types) for the single B/N-doped 11-aGNRs (Figs. 2a/2b). Namely, the single B/N dopant atom behaves as an acceptor/donor impurity in aGNRs, as can be directly visualized in the energy-resolved local density of states (LDOS) plots for the valence band (shaded area in Fig. 2a)/conduction band energy regions (shaded area in Fig. 2b) shown in

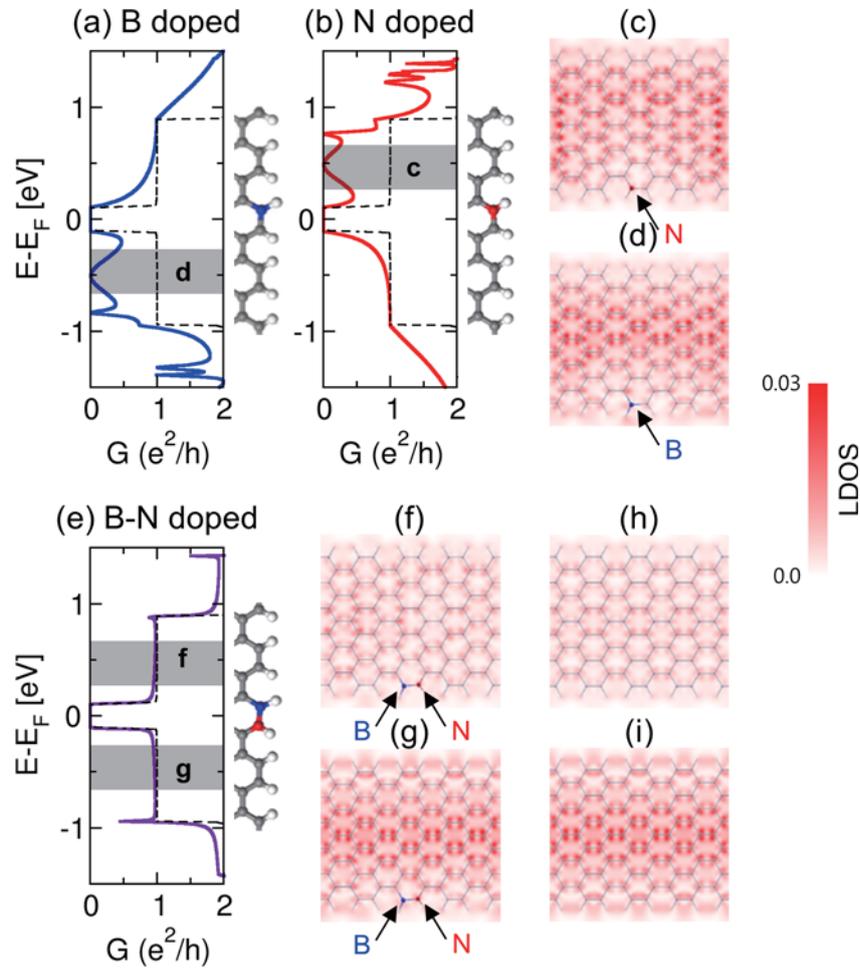

**Fig. 2** Transmissions of (a) single B-doped, (b) single N-doped, (e) B-N-complex doped 11-aGNRs. Transmissions of the pristine 11-aGNRs are shown together (dotted lines). LDOS of the (c) single N-doped, (d) single B-doped, (f,g) B-N-complex doped, and (h,i) pristine 11-aGNRs for the energy ranges $E_F + 0.3$ eV ~ $E_F + 0.7$ eV ((c), (f) and (h)) and $E_F - 0.7$ eV ~ $E_F - 0.3$ eV ((d), (g) and (i)).

Compared with the single B/N doping cases, we find that B-N-complex edge doping almost perfectly recovers the transmission of pristine aGNRs (Fig. 2e) (see also Supplemental Data Fig. S3). This can be easily understood based on the chemical valency of the connected

B and N atoms (three and five, respectively), which should recover that of the $sp^2$ C network due to the compensation effect. The conductance recovery is again well demonstrated in the energy-resolved LDOS plots shown in Figs. 2f and 2g, which almost match those of a pristine GNR shown in Figs. 2h and 2i. Note that experimentally preparing pristine GNRs with atomically smooth edges is still a formidable challenge [2, 15, 20, 21]. In this context, we propose the B-N-complex doping as a practical solution to anneal the edge defects and accordingly realize "effectively" clean and smooth GNR edges. Combined with the energetic consideration presented above, this result should have significant practical implications, e.g, in the lithographic patterning of GNR devices.

We next move on to the case of zGNRs. In Figs. 3a and 3b (See also Supplementary data Fig. S4), as references, we show the transmissions obtained from the single B and single N edge doped 6-zGNRs, respectively. In the ground state of pristine zGNRs, the spins along each zigzag edge are ferromagnetically aligned whereas two edge states are antiferromagnetically (AF) ordered. This ground state presents the degenerate $\alpha$- and $\beta$-spin bands with a bandgap inversely proportional to the ribbon width and the resulting gapped spin-degenerate transmission spectrum (shown as dotted lines in Figs. 3a and 3b) with zero total spin polarization [7]. The spatial and energetic distributions of these spin states can be visualized in the net-spin ($\alpha - \beta$) LDOS plots (Figs. 3d and 3e), which show that the edge-originated spin states change sign as one moves from the valence to conduction bands.

Upon introducing single substitutional B and N dopant atoms into the zGNR edge, in agreement with previous studies [25-31], we obtain the n-type (Fig. 3a) and p-type (Fig. 3b) transport polarities, respectively. This anomalous acceptor-donor transition results from the Coulomb repulsion between the localized zigzag edge states and the dopant atom states [29-31]. The quasi-bound states induced by a B or N atom break the spin degeneracy, but the the robustness and accessibility of spin polarization is limited in that broad and deep conductance drops appear in both spin channels. This results from the spatially rather long-ranged nature of the defect states, which destroy even the spin currents along the opposite undoped GNR edges (Figs. 3f and 3g).

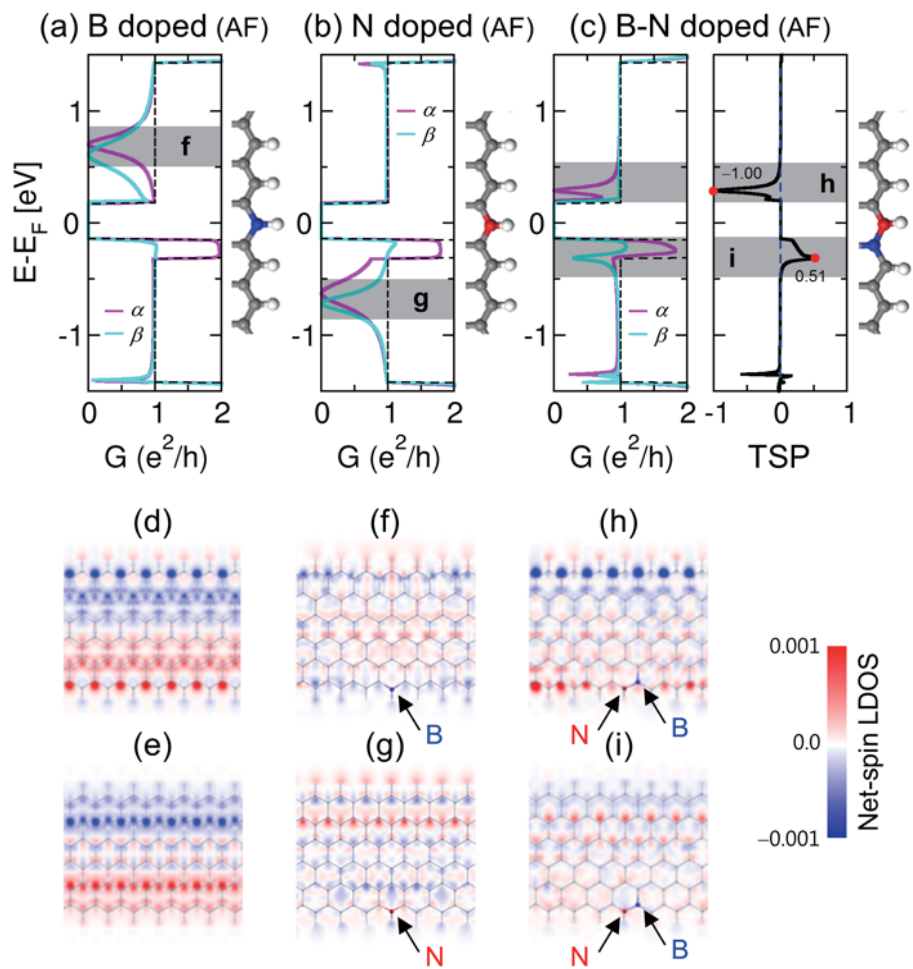

**Fig. 3** Transmissions of (a) single B-doped, (b) single N-doped, and (c) B-N-complex doped 6-zGNRs for the AF spin configurations. Transmissions of pristine 6-zGNR are shown together (dotted lines). In (c), TSP values are also shown. Net-spin LDOS of (f) single B-doped ($E_F + 0.5$ eV ~ $E_F + 0.85$ eV) and (g) single N-doped ($E_F - 0.85$ eV ~ $E_F - 0.5$ eV) 6-zGNRs can be compared with those of (d,e) pristine and (h,i) B-N-complex-doped 6-zGNRs ($E_F + 0.2$ eV ~ $E_F + 0.55$ eV for (d,h) and $E_F - 0.45$ eV ~ $E_F - 0.15$ eV for (e,i)).

The above analysis of single B/N-doping cases suggests that localizing the doped-side edge defect states could be a route to generate more reliable spin-polarized currents. We now show that the B-N complex doping can achieve such a goal. The starting point of this argument is the B-N codoping-induced recovery of conductance (Fig. 3c) established above for aGNRs (see also Supplemental Data Fig. S4). We find that the conductance dips around $E_F + 0.67$ eV (Fig. 3a) and $E_F - 0.67$ eV (Fig. 3b) for the B-doped and N-doped 6-zGNRs, respectively, are restored back to the maximum quantum conductance of 1 $G_0$ ($= e^2/h$). The crucial point is that the conductance recovery is slightly yet systematically incomplete for the spin channel spatially along the B-N doped GNR edge and energetically near the valence and conduction band-edge regions ($\sim E_F - 0.19$ eV and $\sim E_F + 0.29$ eV, respectively). Calculating the transport spin polarization (TSP) defined as

$$TSP(E) = \frac{T_\alpha(E) - T_\beta(E)}{T_\alpha(E) + T_\beta(E)} \qquad (3)$$

we particularly obtain TSP values of −100% in the conduction band edge region. The origin of the sharp conductance dips appearing only for one spin channel in the valence and conduction band edges is indeed the spatially localized nature of B-N-complex defect states (spin currents along the undoped GNR edges are not affected), as can be visualized in the net-spin LDOS plots (Figs. 3h and 3i).

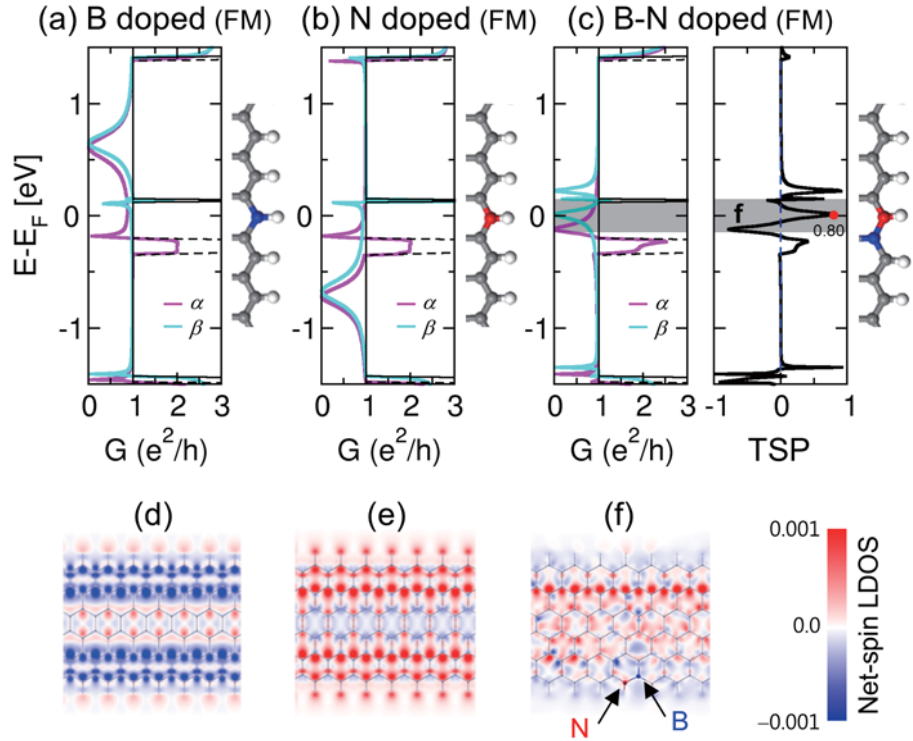

**Fig. 4** Transmissions of (a) single B-doped, (b) single N-doped, and (c) B-N-complex-doped 6-zGNRs for FM spin configurations. Transmissions of pristine 6-zGNR are shown together (dotted lines). In (c), TSP values are also shown. Net-spin LDOS of (d,e) pristine ($E_F - 0.45$ eV ~ $E_F$ and $E_F$ ~ $E_F + 0.45$ eV, respectively) and (f) B-N-complex-doped ($E_F - 0.15$ eV ~ $E_F + 0.15$ eV) 6-zGNRs.

In addition to the ground-state AF spin configuration, we have also considered the ferromagnetic (FM) spin configuration, which is energetically only slightly unstable [17] (Supplemental Data Table S1). Specifically, the energy difference was found to be only 0.002 eV per 12 graphene unit-cell 6-zGNR for the B-N-complex doped case. We find that, unlike the single B or single N doping counterparts (Figs. 4a and 4b, respectively), the B-N-complex doping can support 80% spin-polarized currents at $E_F$ (Fig. 4c) (see also Supplementary Data Fig. S5). As in the AF case, these spin-polarized currents are spatially as well as energetically localized (Fig. 4f), which makes them a source of spin-polarized electrons potentially useful in the nanoscale junction geometry.

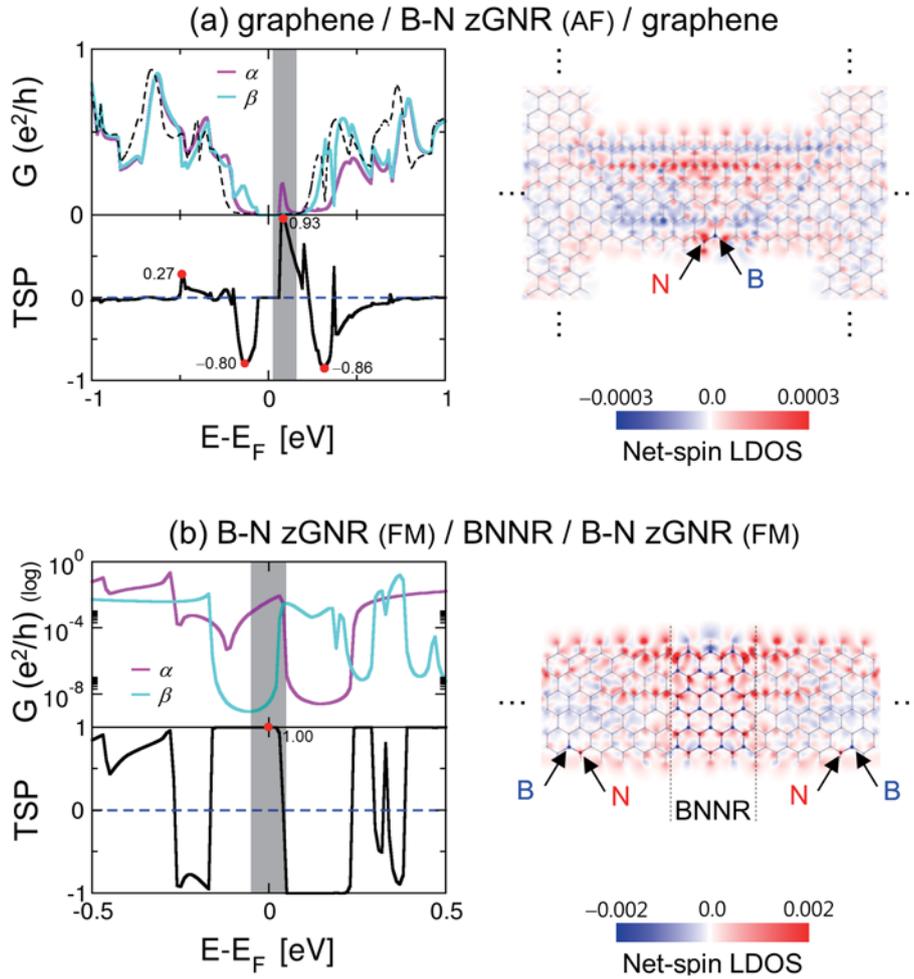

**Fig. 5** (a) Spin-polarized transmissions and TSP obtained for the 11.5 unit-cell ($L$ = 27.05 Å) B-N-doped 6-zGNR (AF) channel sandwiched between graphene electrodes (left panel). Spin-unpolarized transmission from the junction based on a pristine 6-zGNR (AF) channel is shown for comparison (dotted line). Net-spin LDOS for the energy range $E_F$ + 0.05 eV ~ $E_F$ + 0.25 eV (right panel). (b) Spin-polarized transmissions and TSP obtained for the 3.5 unit-cell 6-BNNR channel ($L$ = 18.70 Å) sandwiched between B-N-complex doped 6-zGNR (FM) electrodes (left panel). Net-spin LDOS for the energy range $E_F$ − 0.05 eV ~ $E_F$ + 0.05 eV (right panel).

We have confirmed the robustness of the above-described spin-polarized currents with respect to the variations in the GNR width and doping concentration (see Supplemental Data Figs. S6 and S7). Moreover, we also confirmed that the spin polarized currents survive in the heterojunction geometries, and here we describe two examples. In the first model (Fig. 5a), a B-N-complex edge-doped 6-zGNR in the AF spin configuration was sandwiched between two infinite graphene electrodes. Here, we found spin-polarized conductance with the TSP values of +90% and −80% in the conduction and valence band edges, respectively. In the second junction model (Fig. 5b), between two B-N-complex doped 6-zGNR electrodes in the FM spin configuration, we have inserted a hydrogen passivated boron nitride nanoribbon (BNNR) channel that has the same atomic configuration as 6-zGNRs and is an insulator with a wide bandgap of ~ 4 eV. Again, we found that the spin-polarized currents injected from the B-N-complex doped zGNR electrodes tunnel through the BNNR channel with the TSP value of +100% at $E_F$. For both cases, we can confirm in their LDOS that a fully spin-polarized channel appears in spite of the contacts with graphene electrodes or the BNNR channel.

## 4. Conclusions

In conclusion, based on first-principles calculations, we showed that the B-N-complex GNR edge doping is not only energetically favorable, but also can induce desirable and interesting quantum charge and spin transport properties. We predicted that it can generally recover the excellent charge transport capacity of pristine GNRs. In particular, for the zGNRs, the spatially localized nature of B-N defect states was shown to support robust and tunable spin-polarized currents. For the ground antiferromagnetic state, nearly 100% spin-polarized currents were predicted to flow along the B-N undoped edge near the conduction band-edge region. For the energetically comparable ferromagnetic spin configuration, fully spin-polarized currents along the B-N undoped edge were found to arise near $E_F$. It was confirmed that such spin-polarized states are robust with respect to the GNR width, doping concentration, and heterojunction geometries (both as an AF channel and FM electrodes). With the rapid progress in synthesis techniques [2, 14, 15, 19-21], GNRs have emerged as a promising ingredient of next-generation nanoelectronic and spintronic applications. The proposed B-N-complex edge doping may aid the fabrication of high-quality GNRs and benefit the design of graphene-based spintronic devices.

**Acknowledgements**

This research was supported by Global Frontier Program (No. 2013M3A6B1078881), Basic Science Research Grant (No. 2012R1A1A2044793), and Nano-Material Technology Development Program (No. 2012M3A7B4049888) of the National Research Foundation funded by the Ministry of Science, ICT and Future Planning of Korea. Computational resources were provided by the KISTI Supercomputing Center (No. KSC-2013-C3-046).